**Estimating Covariate-balanced Survival Curve in Distributed Data Environment using Data Collaboration Quasi-Experiment**

*Akihiro Toyoda* [a], *Yuji Kawamata* [b], *Tomoru Nakayama* [c], *Akira Imakura* [d], *Tetsuya Sakurai* [e], *Yukihiko Okada* [f]

[a] Graduate School of Science and Technology, University of Tsukuba, Japan.
toyoda.akihiro.as@alumni.tsukuba.ac.jp, 0009-0008-0427-2587

[b] Center for Artificial Intelligence Research, Tsukuba Institute for Advanced Research, University of Tsukuba, Japan.
yjkawamata@gmail.com, 0000-0003-3951-639X

[c] Graduate School of Science and Technology, University of Tsukuba, Japan.
nakayama.tomoru.tkb_en@u.tsukuba.ac.jp

[d] Institute of Systems and Information Engineering, University of Tsukuba, Tsukuba, Japan. / Center for Artificial Intelligence Research, Tsukuba Institute for Advanced Research, University of Tsukuba, Tsukuba, Japan.
imakura@cs.tsukuba.ac.jp, 0000-0003-4994-2499

[e] Institute of Systems and Information Engineering, University of Tsukuba, Tsukuba, Japan. / Center for Artificial Intelligence Research, Tsukuba Institute for Advanced Research, University of Tsukuba, Tsukuba, Japan.
sakurai@cs.tsukuba.ac.jp, 0000-0002-5789-7547

[f] Institute of Systems and Information Engineering, University of Tsukuba, Tsukuba, Japan. / Center for Artificial Intelligence Research, Tsukuba Institute for Advanced Research, University of Tsukuba, Tsukuba, Japan.
okayu@sk.tsukuba.ac.jp, 0000-0003-4903-4191

**Corresponding Author**: Yuji Kawamata,
yjkawamata@gmail.com, 1-1-1 Tennodai, Tsukuba, Ibaraki 305-8573, Japan

**Abstract**

The sharing of patient-level data necessary for covariate-adjusted survival analysis between medical institutions is difficult due to privacy protection restrictions. We propose a privacy-preserving framework that estimates balanced Kaplan–Meier curves from distributed observational data without exchanging raw data. Each institution sends only the low-dimensional representation obtained through dimensionality reduction of the covariate matrix. Analysts reconstruct the aggregated dataset, perform propensity score matching, and estimate survival curves. Experiments using simulation datasets and five publicly available medical datasets showed that the proposed method consistently outperformed single-site analyses. This method can handle both horizontal and vertical data distribution scenarios and enables the collaborative acquisition of reliable survival curves with minimal communication and no disclosure of raw data.

## Introduction

In medical research, survival outcomes are one of the most important factors for evaluating treatment effects. When estimating treatment effects for survival outcomes from observational data such as electronic medical records, confounding bias is unavoidable because treatment assignment is not random[1,2]. To mitigate such biases and ensure the validity of results, it is essential to appropriately adjust for covariates related to both treatment assignment and outcomes. Specific adjustment methods include multivariate regression, propensity score matching, inverse probability weighting, and stratification, which are standardly used[3,4], and these methods have been shown to reduce confounding bias in observational data[5,6].

Such methods require enough cases and diverse covariates to effectively adjust for confounding, as demonstrated in simulation studies [7–9]. However, data from a single site may suffer from insufficient case numbers and biased patient characteristics, which reduce estimation accuracy and limit external validity. Therefore, integrating data distributed across multiple sites, countries, and regions to analyze them as a large cohort has become essential for improving the reliability of confounding adjustment[10–14]. On the other hand, sharing raw data is challenging due to privacy concerns and regulations. Therefore, establishing a framework that allows for the analysis of multi-site data while maintaining privacy is an urgent task.

Representative methods in survival time analysis include the Cox proportional hazards model[15] and the Kaplan–Meier (KM) curve[16]. Since the Cox model is a multivariate regression framework, bias can be adjusted in a model-based manner by simultaneously incorporating covariates. In recent years, numerous frameworks have been proposed for estimating the Cox model from distributed data while maintaining privacy. First, few-shot algorithms have been proposed where each facility sends only minimal statistics such as event counts, partial likelihood gradients, and Hessian to the server, and parameters are estimated centrally[17,18]. Additionally, a one-shot algorithm has been proposed where each facility reduces the dimension of the covariate matrix and shares it, and the server integrates and redistributes the parameters, ultimately enabling each institution to calculate local parameters[19]. The advantage of these algorithms is that they can adjust the effects of covariates using only the model, without using external indicators such as propensity scores. In addition, a method has been proposed that uses information from each institution to estimate propensity scores, applies inverse probability weighting or matching, and then shares weighted or matched aggregate statistics to estimate hazard ratios[20,21]. These studies enhance the robustness of the model by combining propensity score-based confounding adjustment with Cox model estimation. Such research suggests that the requirement to simultaneously achieve covariate adjustment and Cox model estimation under privacy protection is being met to a certain extent.

While the Cox model performs regression analysis, the KM curve visually represents survival probability, making it an intuitive indicator for medical professionals to understand[22,23]. Major reporting guidelines also recommend presenting KM curves adjusted for covariate[24]. Under privacy protection, it is necessary to estimate KM curves without sharing raw data. Previously proposed methods have been limited to estimating unadjusted KM curves with differential privacy applied to distributed data[25,26], leaving confounding bias unaddressed. As a result, their practicality is limited from the perspectives of clinical decision-making and guideline compliance. In other words, there is still no method available to estimate covariate-balanced KM curves, which are highly valued in clinical settings, while protecting privacy using distributed data.

To address this gap, this study focuses on data collaboration quasi-experiments (DC-QE)[27], one of the distributed data analysis methods. DC-QE is attracting attention as a one-shot framework that enables covariate adjustment and causal effect estimation without sharing raw data. In this method, each institution shares privacy protected representations instead of raw data to perform propensity score analysis. While DC-QE has been shown to be effective for estimating the average treatment effect for binary and continuous outcomes, it is not designed for survival outcomes.

In this study, we propose a new method for estimating covariate-balanced KM curves under privacy protection. The proposed method extends the existing DC-QE framework for privacy preserving quasi-experiments to survival outcomes, enabling collaboration without directly sharing distributed samples or covariates. The main contributions of this study are as follows.

- We proposed the first privacy-preserving framework for estimating covariate-balanced KM curves based on propensity scores in a distributed data environment.
- Through collaboration across the distributed samples, the proposed method identified appropriately matched pairs between the treatment and control groups. Furthermore, through collaboration across the distributed covariates, bias can be effectively reduced, and more accurate survival curves can be estimated.
- Numerical simulation results using an artificial dataset and five open medical datasets show that the proposed method outperforms the analysis performed on the local dataset alone.

# Method

## Data collaboration quasi-experiment for survival outcomes

The algorithm proposed in this study consists of two stages: data aggregation and estimation of covariate-balanced survival curves. In the first phase, collaborative representations are created by analysts from dimension-reduced site-specific data and publicly shared anchor data. In the second stage, covariate-balanced survival curves are estimated using propensity score matching (PSM) and Kaplan-Meier estimation from these representations. Figure 1 shows an overview of the proposed algorithm.

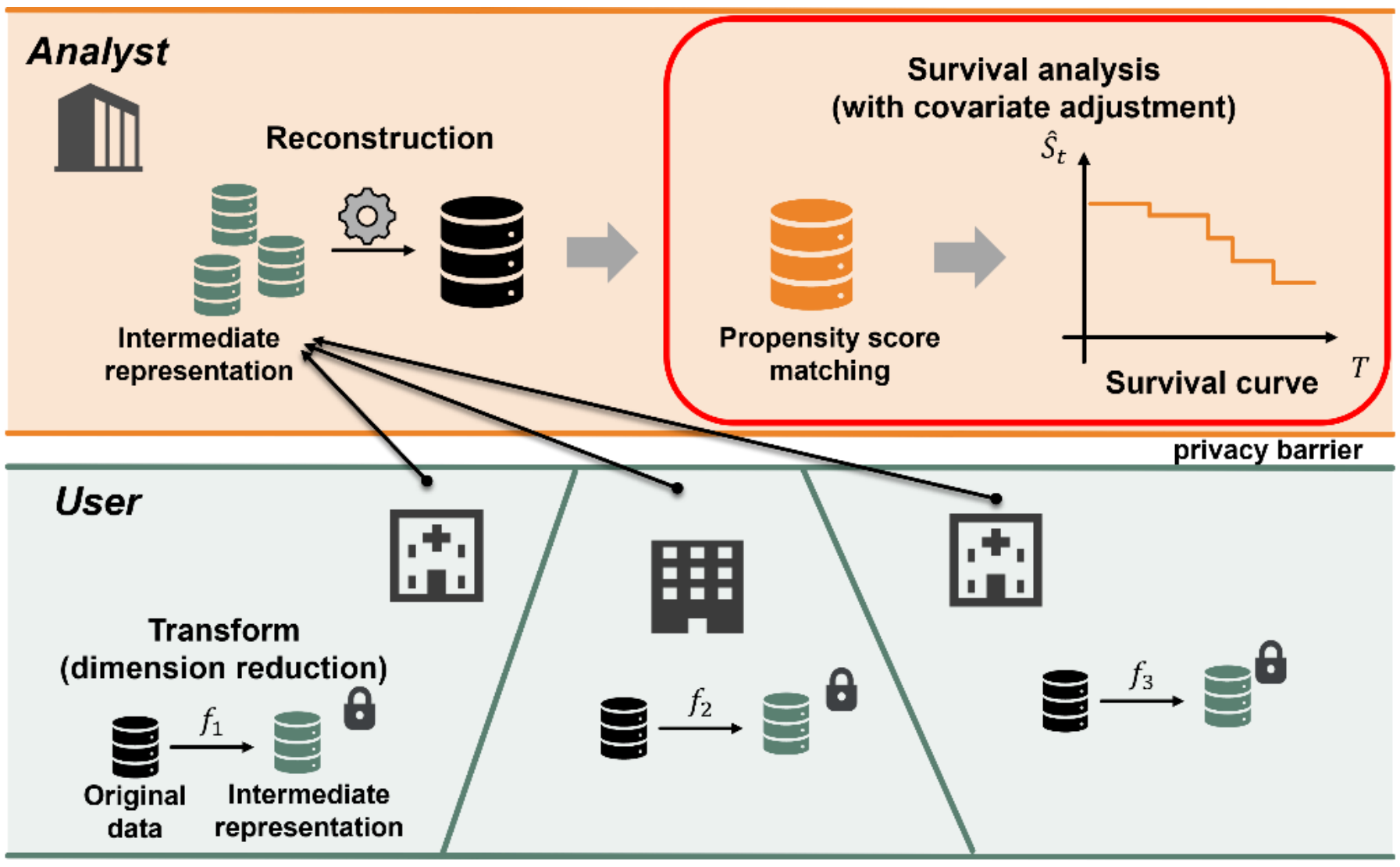

**Figure 1** Outline of the proposed method. This method obtains privacy-preserving survival analysis for distributed data without iterative cross-sites communications.

**Problem settings**

In this paper, we define two roles participating in DC-QE: analysts and users. Users provide data, while analysts are responsible for processing and analyzing aggregated information. Each user $k$ $(k = 1, \dots, c)$ has a private covariate matrix $X_k \in \mathbb{R}^{n_k \times m}$, an observed time vector $\boldsymbol{t}_k \in \mathbb{R}^{n_k}$, an event indicator vector $\boldsymbol{\delta}_k \in \{0,1\}^{n_k}$, and a binary treatment variable $\boldsymbol{Z}_k \in \{0,1\}^{n_k}$ indicating the treatment group and control group. Here, $n$ denotes the sample size and $m$ denotes the number of covariates. Furthermore, we assume that the covariate dataset is managed by $c$ institutions and divided horizontally, and is also divided vertically into $d$ parties with different covariates due to physical or operational reasons. Therefore, the covariate matrix $X$ is divided as follows:

$$X = \begin{bmatrix} X_{1,1} & X_{1,2} & \cdots & X_{1,d} \\ X_{2,1} & X_{2,2} & \cdots & X_{2,d} \\ \vdots & \vdots & \ddots & \vdots \\ X_{c,1} & X_{c,2} & \cdots & X_{c,d} \end{bmatrix}. \quad (1)$$

The $(k, l)$-th party holds a dataset $X_{k,l} \in \mathbb{R}^{n_k \times m_l}$ where $n = \sum_{k=1}^{c} n_k$ and $m = \sum_{l=1}^{d} m_l$.

The estimation targets are the covariate-balanced survival function $S^1(t)$ and $S^0(t)$ for each of the treatment and control group. Under such a distributed setting, it is expected that each party individually will face bias and error in estimation results due to a lack of covariates and insufficient sample size. Furthermore, although the most accurate analysis can be performed by aggregating data from all parties, this is difficult to implement due to restrictions and mechanisms related to privacy protection.

**Phase 1: Construction of Collaboration Representation**

DC-QE adopts the data collaboration framework[28] for data aggregation. In data collaboration, the collaboration representation used for analysis is constructed from dimensionally reduced intermediate representations.

First, each user generates and shares an anchor dataset $X^{\mathrm{anc}} \in \mathbb{R}^{r\times m}$, where $r$ is the number of samples in the anchor dataset. The anchor dataset is a publicly shareable dataset composed of either publicly available or randomly generated dummy data. This is used to compute the reconstruction function $g_k$ (described below). Next, each user constructs an intermediate representation, as follows:

$$\tilde{X}_{k,l} = f_{k,l}(X_{k,l}) \in \mathbb{R}^{n_k \times \tilde{m}_{k,l}}, \tag{2}$$

$$\tilde{X}_{k,l}^{\mathrm{anc}} = f_{k,l}(X_{:,l}^{\mathrm{anc}}) \in \mathbb{R}^{r \times \tilde{m}_{k,l}}, \tag{3}$$

where $f_{k,l}$ is a row-wise linear or nonlinear dimensionality reduction function applied to the local covariate block $X_{k,l}$ at $(k,l)$-th party. In this study, as a concrete example, $f_{k,l}$ is implemented by applying principal component analysis (PCA) to $X_{k,l}$ and retaining the first $\tilde{m}_{k,l}$ principal component scores for each individual; thus $\tilde{X}_{k,l}$ consists of these PCA scores. Other dimensionality reduction methods, such as locality preserving projections or nonnegative matrix factorization, can also be used within the same framework. Here, $\tilde{m}_{k,l}$ denotes the reduced dimensions. We use $X_{:,l}$ to denote the collection of the $l$-th covariate block across all institutions. In other words, $X_{:,l}$ corresponds to the same set of variables (for example, the same laboratory measurements) observed at multiple institutions and stored locally as $X_{k,l}$ at each site. Function $f_{k,l}$ is kept private for each user and differs across users. Through this process, each party generates an intermediate representation, enabling it to provide only low-dimensional representations to analysts without directly disclosing its private data. Subsequently, each user shares only the intermediate representations $\tilde{X}_{k,l}$ and $\tilde{X}_{k,l}^{\mathrm{anc}}$, with the analyst. To handle survival outcomes and treatment indicator, each party additionally shares $t_i$ , $\delta_i$ and $Z_i$. Importantly, the original covariate data $X_{k,l}$ were not directly shared.

The analyst then constructs a collaboration representation from shared intermediate representations. Each party's intermediate representation $\tilde{X}_k$ is expressed in its own local coordinate system, so even the anchor representations $\tilde{X}_k^{\mathrm{anc}}$are not aligned across institutions. The remedy is to map all intermediate representations into a common coordinate system. Concretely, if we construct a shared coordinate matrix $H$ and linear mapping functions $g_k$ such that $g_k(\tilde{X}_k^{\mathrm{anc}}) \approx H$ for the anchor data, then each party's space can be aligned to $H$ and can be integrated. The singular value decomposition (SVD) is employed to derive $H$, following Imakura et al. (2020)[28]. The details are described below.

The goal is to find linear mapping function $g_k$ such that the projected anchor datasets satisfy $g_k(\tilde{X}_k^{\mathrm{anc}}) \approx g_{k'}(\tilde{X}_{k'}^{\mathrm{anc}})$ for $k \neq k'$. Assuming that the mapping function $g_k$ from intermediate representations to collaboration representations is a linear transformation function, we write

$$\check{X}_k^{\mathrm{anc}} = g_k(\tilde{X}_k) = \tilde{X}_k G_k, \qquad \check{X}_k^{\mathrm{anc}} = g_k(\tilde{X}_k^{\mathrm{anc}}) = \tilde{X}_k^{\mathrm{anc}} G_k, \tag{4}$$

where $G_k \in \mathbb{R}^{\tilde{m}_k \times \tilde{m}}$ is the matrix representation of $g_k$, and $\tilde{X}_k = [\tilde{X}_{k,1}, \tilde{X}_{k,2}, \ldots, \tilde{X}_{k,d}]$. These transformations can be determined by solving a least-squares problem:

$$\min_{G_1,\ldots,G_c,H} \sum_{k=1}^{c} \left\| H - \tilde{X}_k^{\mathrm{anc}} G_k \right\|_F^2 . \tag{5}$$

This problem is difficult to solve directly. However, an approximate solution can be derived using SVD. We compute a low-rank SVD

$$[\tilde{X}_1^{\mathrm{anc}}, \tilde{X}_2^{\mathrm{anc}}, \ldots, \tilde{X}_c^{\mathrm{anc}}] \approx U_1 \Sigma_1 V_1^T, \tag{6}$$

where $\Sigma_1 \in \mathbb{R}^{\check{m}\times\check{m}}$ is a diagonal matrix containing the largest singular values, and $U_1$ and $V_1$ are column-orthogonal matrices corresponding to the left and right singular vectors, respectively. We set the shared coordinate matrix as $H = U_1$. Then each transformation matrix $G_k$ is calculated as follows:

$$G_k = \left(\tilde{X}_k^{\mathrm{anc}}\right)^{\dagger} U_1, \tag{7}$$

where † denotes the Moore–Penrose pseudoinverse. By construction, the transformed anchor data satisfy

$$\check{X}_k^{\mathrm{anc}} = \tilde{X}_k^{\mathrm{anc}} G_k \approx H \approx \tilde{X}_{k'}^{\mathrm{anc}} = \check{X}_{k'}^{\mathrm{anc}}, \tag{8}$$

which means that the anchor data from all organizations are aligned in the common collaboration space. Using these transformations, the final collaboration representation $\check{X}$ is given by:

$$\check{X} = \begin{bmatrix} \check{X}_1 \\ \check{X}_2 \\ \vdots \\ \check{X}_c \end{bmatrix} \in \mathbb{R}^{n\times\check{m}}. \tag{9}$$

**Phase 2: Estimation of Covariate-balanced Survival Curves**

In the second phase, the analyst performs PSM and survival curve estimation from $\check{X}, t, \delta$ and Z. The propensity score on the collaboration representation is defined as

$$\alpha_i = \Pr(z_i = 1 | \check{x}_i), \tag{10}$$

where $\check{x}_i$ denotes the value of the $i$-th patient in the collaboration representation $\check{X}$. Using methods such as logistic regression, the analyst can estimate $\alpha_i$ from the aggregated $\check{X}$ and $Z$ obtained during the first phase. This enables the estimation of individual propensity scores without sharing raw private data. Now that the propensity scores are calculated, we can make a dataset by matching samples with similar propensity scores in the treatment and control groups using PSM. From this dataset, we can estimate the survival curves for each group using the Kaplan–Meier estimator[16].

Through these steps, our proposed method enables the integration of information from multiple parties and the estimation of treatment effects on survival outcomes while preserving privacy. Although the proposed method is based on the DC-QE algorithm, it differs because the outcome of interest is the survival curve. Consequently, survival time data were included among the shared information, and Kaplan–Meier estimation was employed for treatment effect estimation. The pseudocode for the proposed method is presented in Algorithm 1.

| Algorithm 1: Proposed method. | |
|---|---|
| | **Input:** covariate dataset $X$, treatment $Z$, time $t$, event $\delta$ |
| | **Output:** estimated survival curve |
| I. | Construction of collaboration representation |
| | User side $(k,l)$ |
| 1: | Generate anchor dataset $X_{k,l}^{\text{anc}}$ and share with all users |
| 2: | Set $X_{:,l}^{\text{anc}}$ |
| 3: | Generate $f_{k,l}$ and compute $\tilde{X}_{k,l}, \tilde{X}_{k,l}^{\text{anc}}$ |
| 4: | Share $\tilde{X}_{k,l}, \tilde{X}_{k,l}^{\text{anc}}, t_k, \delta_k, Z_k$ to the analyst |
| | Analyst side |
| 5: | Get $\tilde{X}_{k,l}, \tilde{X}_{k,l}^{\text{anc}}, t_k, \delta_k, Z_k$ for all $k, l$ |
| 6: | Compute $g_k$ from $\tilde{X}_{k,l}^{\text{anc}}$ |
| 7: | Compute $\check{X}_k$ for all $k$ |
| II. | Estimation of covariate-balanced survival curve |
| 8: | Set $\check{X}, Z, t, \delta$ |
| 9: | Estimate propensity score $\alpha$ from $\check{X}, Z$ |
| 10: | Estimate survival curve from $\alpha, t, \delta$ using Kaplan–Meier method |

**Privacy Preservation of Proposed Method**

In this section, we describe how privacy and confidentiality are preserved using the proposed method. Similar to prior works on data collaboration analysis[19,29], our approach incorporates two layers of privacy protection to safeguard sensitive data, $X_{k,l}$.

- **First Layer:** Under this protocol, no participant other than the data owner has access to $X_{k,l}$, and transformation function $f_{k,l}$ remains private. Because neither the inputs nor the outputs of $f_{k,l}$ are shared with the other participants, $f_{k,l}$ remains confidential. Consequently, others cannot infer the original data $X_{k,l}$ from the shared intermediate representation $\tilde{X}_{k,l}$.
- **Second Layer:** The second layer of protection stems from the fact that $f_{k,l}$ is a dimensionality reduction function. Even if $f_{k,l}$ were leaked, the original data $X_{k,l}$ could not be reconstructed from $\tilde{X}_{k,l}$ because of the dimensionality reduction properties. This privacy guarantee aligns with the concept of $\varepsilon$-DR (Dimensionality Reduction) privacy[30].

However, it should be noted that some aggregate statistical properties, such as the means or variances of the original data, could potentially leak through the anchor data. This is a potential risk inherent to the data collaboration framework.

## Common Settings and Evaluation Scheme

To validate the effectiveness of the proposed method, three benchmark methods are established for comparison.

a. Central Analysis (CA):

In CA, the analyst has access to the entire dataset, including covariates $X$, treatment assignments $Z$, survival times $t$, and event indicators $\delta$, and estimates survival curves based on this complete information. As there are no data-sharing constraints, CA represents the ideal baseline.

b. Local Analysis (LA):

In LA, each user independently analyzes their own data without sharing any information externally. Each user possesses only their own dataset—covariates $X_k$, treatment $Z_k$, survival times $t_k$, and event indicators $\delta_k$ —and estimates survival curves solely based on this limited information.

c. Local Matching and Central Analysis (LMCA):

Since survival curve estimation requires $t$, $\delta$, and $Z$, under the considered setting, it is possible to perform survival curve estimation by sharing only these variables after local matching. Specifically, each user conducts PSM independently using their own $X_k$ and $Z_k$, then shares only the matched survival times $t_k^{\mathrm{matched}}$, event indicators $\delta_k^{\mathrm{matched}}$, and treatment indicotors $Z_k^{\mathrm{mathced}}$ with the analyst, without sharing the original covariates $X_k$. Thus, privacy is preserved using this method.

In this study, it is desirable that the results of the proposed method are closer to those obtained by CA than to those obtained by LA.

The experimental conditions were as follows: The anchor dataset $X^{\mathrm{anc}}$ was generated using random numbers sampled uniformly within the range defined by the minimum and maximum values of the entire dataset. This type of anchor data was also employed in Imakura and Sakurai[28]. The number of samples in the anchor dataset $r$ was set to be equal to the total number of samples $n$. PCA was applied to each participant as a dimensionality reduction function. PCA is one of the most widely used techniques for reducing dimensionality.

The analyst estimates propensity scores using logistic regression, which is commonly adopted in the literature on propensity scores. Following the approach of Kawamata et al.[27], the logistic regression model used in the experiments consisted of a linear combination of the collaboration representation and an intercept term for the proposed method, whereas for central and local analyses, the model consisted of a linear combination of the covariates and an intercept term. Caliper matching was employed for matching, where the caliper width was set to 0.2 times the standard deviation of the logit of the estimated propensity scores, as recommended by Austin[31].

The performance of each method was evaluated from two perspectives: covariate balance and the accuracy of survival curve estimation. For covariate balance, we used three metrics: the sample size of the matched pairs created through PSM, the inconsistency of the estimated propensity scores, and the balance of covariates after matching. For survival curve estimation, we compared the accuracy of the estimated survival curves across different methods.

**Inconsistency of Propensity Score with CA**

To evaluate the accuracy of the estimated propensity scores, we used the following inconsistency measures:

$$\mathrm{InconsistencywithCA}(\hat{\boldsymbol{e}}, \hat{\boldsymbol{e}}^{\mathrm{CA}}) = \sqrt{\frac{1}{n}\Sigma_{i=1}^{n}\left(\hat{e}_i - \hat{e}_i^{\mathrm{CA}}\right)^2}, \tag{11}$$

where $\hat{\boldsymbol{e}} = [\hat{e}_1, \dots, \hat{e}_n]^T$ denotes the estimated propensity scores, and $\hat{\boldsymbol{e}}^{\mathrm{CA}} = \left[\hat{e}_1^{\mathrm{CA}}, \dots, \hat{e}_n^{\mathrm{CA}}\right]^T$ denotes the propensity scores estimated by CA. A smaller value indicates that the estimated propensity scores were closer to those obtained by CA.

**Covariate Balance**

To evaluate covariate balance improvement through PSM, we used the standardized mean difference (SMD), a commonly adopted metric[32], defined as:

$$d^j = \frac{\bar{x}_T^j - \bar{x}_C^j}{\sqrt{\frac{s_T^j - s_C^j}{2}}}, \tag{12}$$

where $\bar{x}_T^j$ and $s_T^j$ are the mean and standard deviation of covariate $x^j$ in the treatment group, and $\bar{x}_C^j$ and $s_C^j$are those in the control group, respectively. For the overall covariate balance assessment, we employed the maximum absolute standardized mean difference (MASMD), defined as:

$$\mathrm{MASMD}(d) = \max_j \left|d^j\right|, \tag{13}$$

where $\boldsymbol{d} = [d^1, \dots, d^m]$. Similar to the SMD, the MASMD measures the bias in covariate distributions between the treatment and control groups, with smaller values indicating better covariate balance.

**Accuracy of Estimated Survival Curve**

To assess how faithfully each distributed or local method reproduces the survival curves obtained by central analysis (CA), we evaluated the closeness between the estimated survival curves and the CA-estimated curves. Here, CA is treated as a realistic benchmark corresponding to conventional centralized propensity score matching analysis. It should be emphasized that CA itself is an estimator, not the true survival function. Therefore, the "gap" metric described later specifically quantifies proximity to CA, while comparison to the true survival function is separately evaluated using metrics based on RMST. Let $\hat{S}_{m,z}(t)$ denote the estimated survival curve for treatment group $z \in \{0,1\}$ obtained by method $m$, and let $\hat{S}_{\mathrm{CA},z}(t)$ be the corresponding curve from the central analysis.

The gap between the survival curves of method $m$ and CA for group $z$ was defined as:

$$\mathrm{Gap}_{\mathrm{CA}}^{(m,z)} = \sqrt{\frac{1}{T}\sum_{i=1}^{T}\left(\hat{S}_{m,z}(t_i) - \hat{S}_{\mathrm{CA},z}(t_i)\right)^2}, \tag{14}$$

where $\hat{S}_{m,z}(t) = [\hat{S}_{m,z}(t_0), \dots, \hat{S}_{\mathrm{CA},z}(t_T)]$. A smaller Gap value indicates that the survival curve estimated by method m approximates the survival curve derived from the central analysis of the corresponding treatment group. In real-world data experiments, this CA-based Gap serves as a direct measure of how well each distributional method reproduces the covariate-balanced survival curve obtained under centralized PSM.

**Metrics based on restricted mean survival time (RMST)**

RMST provides not only a numerical summary of survival curves but also a clear definition of the effect we aim to estimate: the average survival time up to a clinically relevant time point $\tau$ in each treatment group and their difference. For a prespecified truncation time $\tau > 0$, the restricted mean survival time (RMST) for group $z$ under method $m$ is defined as

$$\hat{\mu}_{m,z} = \int_0^{\tau} \hat{S}_{m,z}(t)\, dt. \tag{15}$$

The corresponding RMST difference for method $m$ is

$$\widehat{\Delta}_m(\tau) = \hat{\mu}_{m,1}(\tau) - \hat{\mu}_{m,0}(\tau). \tag{16}$$

We use this quantity to define two additional performance measures: (i) the agreement with the central analysis, and (ii) the bias with respect to the true survival functions in the simulation study.

1. Difference of RMST difference relative to the central analysis:
   Using the RMST difference obtained by the central propensity score matching analysis (CA) $\widehat{\Delta}_{\mathrm{CA}}(\tau)$, we quantify how closely each comparison method approximates the estimated results as follows:
   $$D_m^{\mathrm{CA}}(\tau) = \left|\widehat{\Delta}_m(\tau) - \widehat{\Delta}_{\mathrm{CA}}(\tau)\right|. \tag{17}$$
   A value of $D_m^{\mathrm{CA}}(\tau)$ close to zero indicates that method $m$ yields RMST differences with covariate balance comparable to that of central analysis. In real world data experiments, since the true survival function is unknown, only comparisons based on this CA are available.
2. Difference of RMST difference relative to the true survival function (simulation study only):
   In synthetic data experiments, the true covariate-balanced survival function $S_z^{\mathrm{true}}(t)$ can be defined from the data generation process for $z \in \{0,1\}$ (see the Method Section). The corresponding true RMST and its difference are as follows:
   $$\mu_z^{\mathrm{true}}(\tau) = \int_0^{\tau} S_z^{\mathrm{true}}(t)\, dt, \qquad \Delta^{\mathrm{true}}(\tau) = \mu_1^{\mathrm{true}}(\tau) - \mu_0^{\mathrm{true}}(\tau). \tag{18}$$
   For each method $m$, we define the difference of RMST difference relative to the true effect as follows:

$$D_m^{\text{true}}(\tau) = \left|\widehat{\Delta}_m(\tau) - \Delta^{\text{true}}(\tau)\right|. \tag{19}$$

This value $D_m^{\text{true}}(\tau)$ quantity directly measures how accurately method $m$ recovers the difference in true RMST suggested by the data generation model.

As with other metrics, we report the mean and standard deviation across repeated trials. A value closer to zero indicates smaller bias and less variability, suggesting the estimator is more stable across simulation datasets.

For all RMST-based metrics, the truncation time $\tau$ was determined from the data in both synthetic and real data analyses. Specifically, for each scenario or dataset, the minimum value of the maximum observed time in the treatment and control groups was set as $\tau$, using all samples prior to resampling or matching.

All numerical experiments were conducted on a Windows 11 machine equipped with an Intel(R) Core(TM) i7-1255U @ 1.70 GHz processor and 16 GB RAM, using Python 3.8.

## Experiment I: Synthetic Data

**Simulation Settings**

We conducted validation experiments using synthetic data. The synthetic dataset contained 1,000 samples, each with six covariates, simulated baseline covariates, treatment assignments, and event times. Covariates $\boldsymbol{x}_i = \left[x_i^1, \dots, x_i^6\right]$ are generated from a multivariate normal distribution.

$$\boldsymbol{x}_i \sim \mathcal{N}(0, S), \qquad (i = 1, \dots, 1{,}000), \tag{20}$$

where $\mathcal{N}(0, S)$ denotes a normal distribution with mean zero, and the covariance matrix $S$ is given by

$$S = \begin{bmatrix} 1 & 0.5 & 0.5 & 0 & 0 & 0 \\ 0.5 & 1 & 0.5 & 0 & 0 & 0 \\ 0.5 & 0.5 & 1 & 0 & 0 & 0 \\ 0 & 0 & 0 & 1 & 0.5 & 0.5 \\ 0 & 0 & 0 & 0.5 & 1 & 0.5 \\ 0 & 0 & 0 & 0.5 & 0.5 & 1 \end{bmatrix}. \tag{21}$$

The probability that a patient $i$ receives treatment ($z_i = 1$) is given by:

$$\Pr(z_i = 1 | x_i) = \frac{1}{1 + \left(\exp\left(\Sigma_{j=1}^6 - \frac{1}{3} x_i^j\right)\right)}. \tag{22}$$

Event times (survival outcomes) were simulated based on the method proposed in Bender et al.[33], using a Weibull distribution. Specifically, survival times were generated according to

$$T_i \sim \left(-\frac{\log\left(\text{uniform}(0,1)\right)}{\lambda \exp\left(\Sigma_{j=1}^6 - \frac{1}{3} x_i + \gamma Z\right)}\right)^{\frac{1}{v}}, \tag{23}$$

where $\lambda$ is the scale parameter, $v$ is the shape parameter of the Weibull distribution, and $\gamma$ represents the marginal treatment effect. In the experiments, we set $\lambda = 2, v = 2$ and $\gamma = -1$. Thus, the survival times depend on both covariates $X$ and the treatment $Z$.

To introduce right censoring, the censoring time was independently generated from an Exponential distribution

$$C_i \sim \mathrm{Exp}(\lambda_C). \tag{24}$$

In the experiments, we set $\lambda_C = 0.3$. The observed time and event indicator were then defined as

$$T^* = \min(T_i, C_i), \qquad \delta_i = I(T_i \leq C_i)\,. \tag{25}$$

This setup ensures that the treatment assignment depends on covariate $X$ and that the covariates also directly affect survival outcomes. Therefore, the covariates act as confounders when estimating the causal effect of treatment on survival. Without adjusting for these confounders, the estimated survival curves might have been biased.

In addition, for defining the "true" covariate-balanced survival curve and RMST used in the evaluation, we generated potential event time $T_i^{(0)}$ and $T_i^{(1)}$ by fixing $Z_i = 0$ and $Z_i = 1$ in the above Weibull distribution and computed the marginal survival functions

$$S_z^{\mathrm{true}}(t) = \Pr\left(T^{(z)} > t\right), \qquad z \in \{0,1\}, \tag{26}$$

via Monte Carlo simulation (N=100,000) over the covariate distribution. The corresponding true RMST and their difference $\Delta^{\mathrm{true}}(\tau)$ were obtained by numerical integration of $S_z^{\mathrm{true}}(t)$ on $[0, \tau]$.

**How to partition the data**

For data distribution, we assumed horizontal and vertical partitioning with $c = 2$ institutions and $d = 2$ partitions.

$$X = \begin{bmatrix} X_{1,1} & X_{1,2} \\ X_{2,1} & X_{2,2} \end{bmatrix}, X_{i,j} \mathbb{R}^{500\times 5}. \tag{27}$$

The datasets were randomly divided such that each institution had an equal number of samples. Each experiment was repeated 100 times.

In this setting, we compare the results for the following cases: In this simulation dataset, the distribution of covariates does not depend on sample $i$ and the samples are randomly split; to eliminate redundancy in the results, we represent the results of the (1,1)-th user in the individual analysis. Because the same results are expected from left-side and right-side collaboration, as well as from upper-side and lower-side collaboration, left-side collaboration (L-clb), upper-side collaboration (T-clb), and overall collaboration (W-clb) are considered as the three proposed methods. (L-clb, T-clb, and W-clb) are considered as the results of the proposed method (DC-QE), and the left collaboration (L-clb) is considered as the result of LMCA. Here, $\widetilde{m}_{k,l} = 2$ and $\breve{m} = 3$ for left-side collaboration, and $\breve{m} = 6$ for upper and overall collaboration.

## Experiment II: Real-World Medical Data

We further evaluated the performance of the proposed method using five publicly available medical datasets, as listed in Table 1. The datasets included four survival analysis datasets from the survival package in R and a real-world Right Heart Catheterization (rhc) dataset.

**Table 1. Datasets used in Experiment II.**

| Dataset | $n$ | $m$ | Description | Criteria for $Z$ |
|---|---|---|---|---|
| colon | 888 | 13 | Chemotherapy data for stage B/C colon cancer | $Z_i = 1$ if sex is 1(male) |
| lung | 167 | 7 | NCCTG lung catheter dataset | $Z_i = 1$ if sex is 1(male) |
| pbc | 276 | 17 | Mayo Clinic primary biliary cirrhosis dataset | $Z_i = 1$ if age > 60 |
| veteran | 137 | 4 | Veterans' Administration lung cancer study | $Z_i = 1$ if age > 60 |
| rhc | 5,735 | 53 | RHC data for critically ill patients | $Z_i = 1$ if RHC was used |

We evaluated the performance of the proposed method by comparing it with LA, CA, and LMCA. Each dataset was horizontally partitioned into $c = 3$ users ($d = 1$), and the samples were randomly divided such that each user held approximately the same number of samples. Each experiment was repeated $20$ times.

# Results

## Experiment I: Synthetic Data

The results of the CA, LA, LMCA, and the three proposed methods (DC-QE (T-clb, L-clb, and W-clb)) are shown in Table 2. The inconsistency metric quantifies the closeness of the estimated propensity scores to those obtained through CA. Among all the methods, DC-QE (T-clb) achieved the smallest inconsistency at 0.0480, indicating that the estimated propensity scores were extremely close to those from the central analysis. This was followed by DC-QE (W-clb) at 0.0857 and DC-QE (L-clb) at 0.1033. In contrast, LA and LMCA exhibited inconsistency values of approximately 0.17, suggesting that without collaborative techniques, whether through simple local analysis or sharing only matched data, the estimated propensity scores deviated substantially from those obtained via central analysis. The particularly low inconsistencies observed for DC-QE (T-clb) and DC-QE (W-clb) indicate that the proposed method effectively approximates the central analysis by leveraging additional covariate information.

Regarding the covariate balance, as measured using the MASMD metric, CA achieved the lowest value of 0.1211, reflecting the best covariate balance. LA and LMCA exhibited larger MASMD values (0.6820 and 0.6720, respectively), implying that when propensity score estimation and matching were performed independently for each user, covariate balance was not sufficiently achieved. In contrast, the proposed method (DC-QE) consistently achieved smaller MASMD values than LA across all collaboration settings, demonstrating that even in a distributed environment, our method can achieve a covariate balance closer to that of centralized analysis.

The proximity of the estimated survival curve and the CA curve were evaluated for both the treatment group and the control group using the Gap metric. DC-QE (W-clb) achieved the smallest gap in both the treatment and control groups (0.0259 and 0.0275, respectively), followed by DC-QE (L-clb) and DC-QE (T-clb), both of which significantly outperformed LA and LMCA. This demonstrates that the collaborative approach can faithfully reproduce Kaplan-Meier curves obtained from centralized PSM. These results suggest that DC-QE (W-clb) benefits from both an increased number of samples through sample-level collaboration and enhanced covariate information through covariate-level collaboration, resulting in survival curve estimates that closely approximate those of a centralized analysis.

Furthermore, we evaluated how well each method reproduces the RMST difference estimated by CA using the difference of RMST difference relative to CA. Among the distributed methods, DC-QE (W-clb) showed the smallest deviation from the CA (mean 0.0425), while DC-QE (T-clb) and DC-QE (L-clb) achieved very similar values (approximately 0.05). In contrast, LA and LMCA exhibited much larger errors (approximately 0.10–0.11). These results confirm that the proposed DC-QE framework can recover the covariate-balanced estimand from CA with substantially smaller errors than other local methods.

In synthetic data environments, the data generation process defines the true covariate-adjusted RMST difference. Therefore, we calculated the difference of RMST difference relative to the true survival function. Since PSM was performed on finite samples, the estimated CA still exhibited bias (mean 0.0690) relative to the true value. However, all DC-QE variants reduced bias compared to LA and LMCA. Among distributed methods, DC-QE (W-clb) achieved the smallest bias (0.0854), followed closely by DC-QE (T-clb) (0.0886). In contrast, DC-QE (L-clb) exhibited slightly larger bias (0.0998). LA and LMCA exhibited even larger biases (approximately 0.16).

**Table 2. Means and standard errors (in parentheses) of performance measures in Experiment I.**

| Method | Sample size after Matching | MASMD | Inconsistency | Gap(T) | Gap(C) | $D^{\mathrm{CA}}$ | $D^{\mathrm{true}}$ |
|---|---|---|---|---|---|---|---|
| LA | 370.18 (20.71) | 0.6820 (0.1059) | 0.1705 (0.014) | 0.0430 (0.0159) | 0.0547 (0.0196) | 0.1096 (0.0526) | 0.1656 (0.0686) |
| LMCA (L-clb) | **739.84** (28.66) | 0.6720 (0.0791) | 0.1792 (0.0149) | 0.0322 (0.0133) | 0.0477 (0.0149) | 0.1036 (0.0416) | 0.1632 (0.047) |
| DC-QE (T-clb) | 310.16 (21.88) | **0.1847** (0.0508) | **0.0480** (0.0124) | 0.0382 (0.0156) | 0.0396 (0.0145) | 0.0530 (0.0409) | <u>0.0886</u> (0.0609) |
| DC-QE (L-clb) | 690.20<br>(74.26) | 0.3404<br>(0.1538) | 0.1033<br>(0.0518) | 0.0262<br>(0.0137) | 0.0331<br>(0.0152) | 0.0526<br>(0.0408) | 0.0998<br>(0.0576) |
| DC-QE (W-clb) | 668.18<br>(51.51) | 0.2772<br>(0.1210) | 0.0857<br>(0.0404) | **0.0259**<br>(0.0111) | **0.0275**<br>(0.0121) | **0.0425**<br>(0.0315) | **0.0854**<br>(0.0502) |
| CA (reference) | 624.58<br>(29.81) | 0.1211<br>(0.0229) | 0.0000<br>(0.0000) | 0.0000<br>(0.0000) | 0.0000<br>(0.0000) | 0.0000<br>(0.0000) | 0.0690<br>(0.0449) |

(Gap(C): Gap for control group, Gap(T): Gap for treatment group, $D^{\mathrm{CA}}$: Difference of RMST difference relative to CA, $D^{\mathrm{true}}$: Difference of RMST difference relative to the true survival function, LA: local analysis, LMCA: local matching and central analysis, CA: central analysis. **The best result** is bold, and the second-best result is underlined.)

## Experiment II: Real-World Medical Data

We further evaluated the performance of the proposed method using five publicly available medical datasets. Table 3 summarizes the mean and standard deviation of the four evaluation metrics for each dataset (colon, lung, pbc, veteran, and rhc). Across all datasets, the proposed method (DC-QE) consistently demonstrated superior or at least comparable performance relative to LA and LMCA in terms of inconsistency metric, MASMD, and Gap metric. These results suggest that DC-QE can achieve an accuracy close to that of the CA even under distributed data settings.

Specifically, compared to LA, DC-QE achieved substantially smaller MASMD values and inconsistency levels that were close to those of CA. Moreover, for the Gap metric, which measures survival curve estimation accuracy, DC-QE consistently outperformed LA, particularly for the control group (Gap(C)), where the estimated survival curves were often much closer to those obtained by central analysis.

When compared with LMCA, DC-QE generally achieved smaller inconsistencies and Gap values. However, for MASMD, no consistent pattern was observed; in some datasets, LMCA exhibited smaller MASMD values than DC-QE. This indicates that while DC-QE outperforms LMCA in terms of treatment effect estimation accuracy and survival curve estimation, the advantage of covariate balance is dataset-dependent.

Since the true survival function is unknown in the real datasets, we focused on the difference of RMST difference relative to CA. $D^{\mathrm{CA}}$ varies across datasets, likely reflecting differences in sample size, observation period, and confounding strength specific to each real dataset. For multiple datasets (lung, veterans, rhc), DC-QE showed the smallest bias in $D^{\mathrm{CA}}$. In the other two data sets (colon and pbc), LMCA showed a slightly smaller deviation than DC-QE, but DC-QE still significantly improved LA.

**Table 3. Means and standard errors (in parentheses) of performance measures in Experiment**

**II**.

| colon | | MASMD | Inconsistency | Gap(T) | Gap(C) | $D^{CA}$ |
|---|---|---|---|---|---|---|
| | LA | 0.115(0.027) | 0.083(0.021) | 0.028(0.008) | 0.031(0.011) | 37.567(29.292) |
| | LMCA | **0.070**(0.023) | 0.075(0.013) | 0.009(0.003) | 0.009(0.004) | **22.915**(13.898) |
| | DC-QE | 0.073(0.017) | **0.016**(0.008) | **0.008**(0.003) | **0.007**(0.003) | 23.583(11.290) |
| | CA(reference) | 0.048(0.000) | 0.000(0.000) | 0.000(0.000) | 0.000(0.000) | 0.000(0.000) |
| **lung** | | **MASMD** | **Inconsistency** | **Gap(T)** | **Gap(C)** | $D^{CA}$ |
| | LA | 0.311(0.116) | 0.137(0.042) | 0.102(0.040) | 0.106(0.048) | 95.063(58.425) |
| | LMCA | **0.178**(0.060) | 0.133(0.027) | 0.052(0.023) | 0.035(0.016) | 32.585(26.802) |
| | DC-QE | 0.218(0.078) | **0.080**(0.021) | **0.040**(0.016) | **0.021**(0.009) | **22.746**(17.723) |
| | CA(reference) | 0.181(0.000) | 0.000(0.000) | 0.000(0.000) | 0.000(0.000) | 0.000(0.000) |
| **pbc** | | **MASMD** | **Inconsistency** | **Gap(T)** | **Gap(C)** | $D^{CA}$ |
| | LA | 0.625(0.157) | 0.139(0.024) | 0.111(0.044) | 0.164(0.089) | 515.906(377.043) |
| | LMCA | 0.379(0.110) | 0.166(0.033) | **0.049**(0.021) | 0.123(0.050) | **354.231**(231.089) |
| | DC-QE | **0.288**(0.094) | **0.044**(0.012) | **0.049**(0.019) | **0.080**(0.030) | 414.033(170.772) |
| | CA(reference) | 0.243(0.000) | 0.000(0.000) | 0.000(0.000) | 0.000(0.000) | 0.000(0.000) |
| **veteran** | | **MASMD** | **Inconsistency** | **Gap(T)** | **Gap(C)** | $D^{CA}$ |
| | LA | 0.296(0.128) | 0.128(0.038) | 0.117(0.068) | 0.106(0.039) | 43.691(24.374) |
| | LMCA | 0.182(0.083) | 0.115(0.026) | 0.070(0.027) | 0.034(0.014) | 29.725(12.249) |
| | DC-QE | **0.169**(0.069) | **0.057**(0.029) | **0.051**(0.018) | **0.020**(0.008) | **18.715**(12.229) |
| | CA(reference) | 0.243(0.000) | 0.000(0.000) | 0.000(0.000) | 0.000(0.000) | 0.000(0.000) |
| **rhc** | | **MASMD** | **Inconsistency** | **Gap(T)** | **Gap(C)** | $D^{CA}$ |
| | LA | 0.086(0.011) | 0.066(0.006) | 0.018(0.007) | 0.021(0.008) | 22.177(13.679) |
| | LMCA | **0.064**(0.009) | 0.074(0.002) | **0.007**(0.002) | 0.007(0.002) | 6.595(4.420) |
| | DC-QE | 0.073(0.016) | **0.025**(0.008) | **0.007**(0.002) | **0.006**(0.002) | **6.273**(4.062) |
| | CA(reference) | 0.056(0.000) | 0.000(0.000) | 0.000(0.000) | 0.000(0.000) | 0.000(0.000) |

(Gap(C): Gap for control group, Gap(T): Gap for treatment group, $D^{CA}$: RMST difference relative to the central analysis, LA: local analysis, LMCA: local matching and central analysis, CA: central analysis. **The best** results are highlighted.)

Figure 2 shows the survival curves estimated using each method. Overall, the proposed method (DC-QE) yields survival curves that are visually closer to those obtained by the CA than to those obtained by LA and LMCA. This tendency was particularly evident in datasets with relatively small sample sizes, such as the lung, pbc, and veteran datasets. In these datasets, although LA and LMCA show noticeable deviations from the survival curves obtained using CA, the proposed method can approximate the central analysis survival curves with higher accuracy.

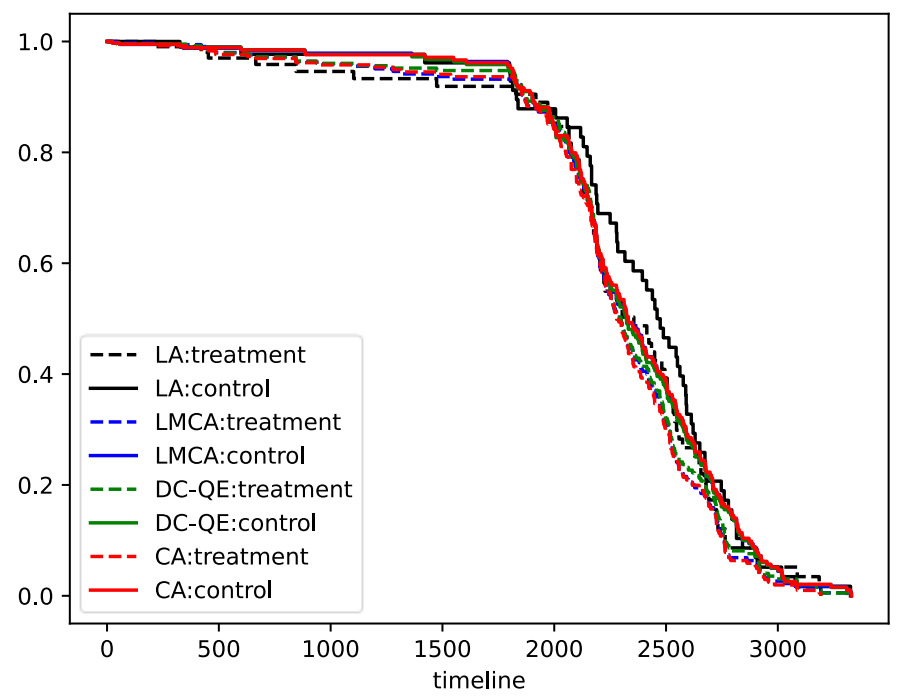

(a) colon

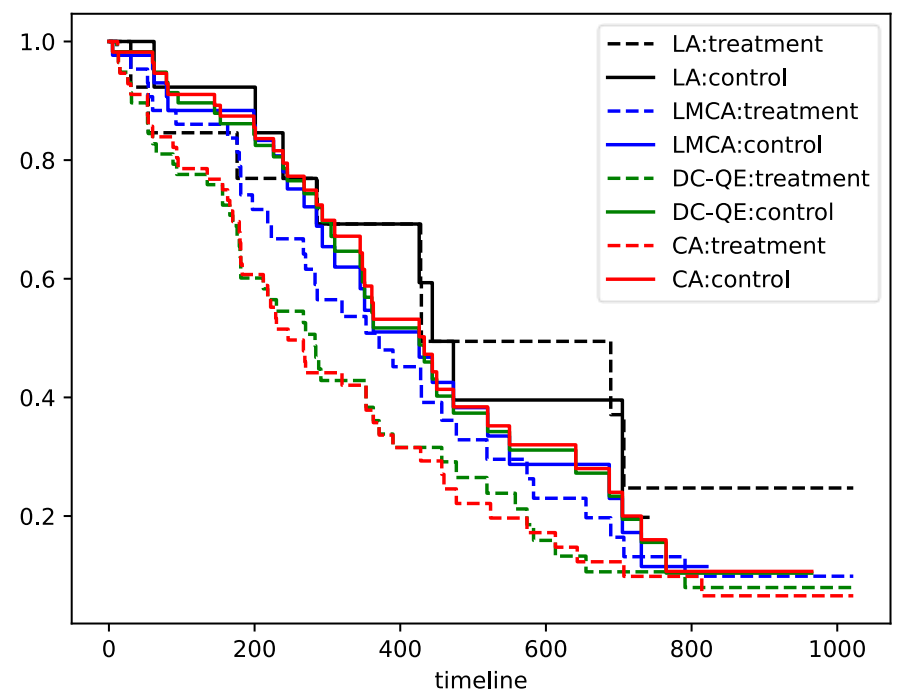

(b) lung

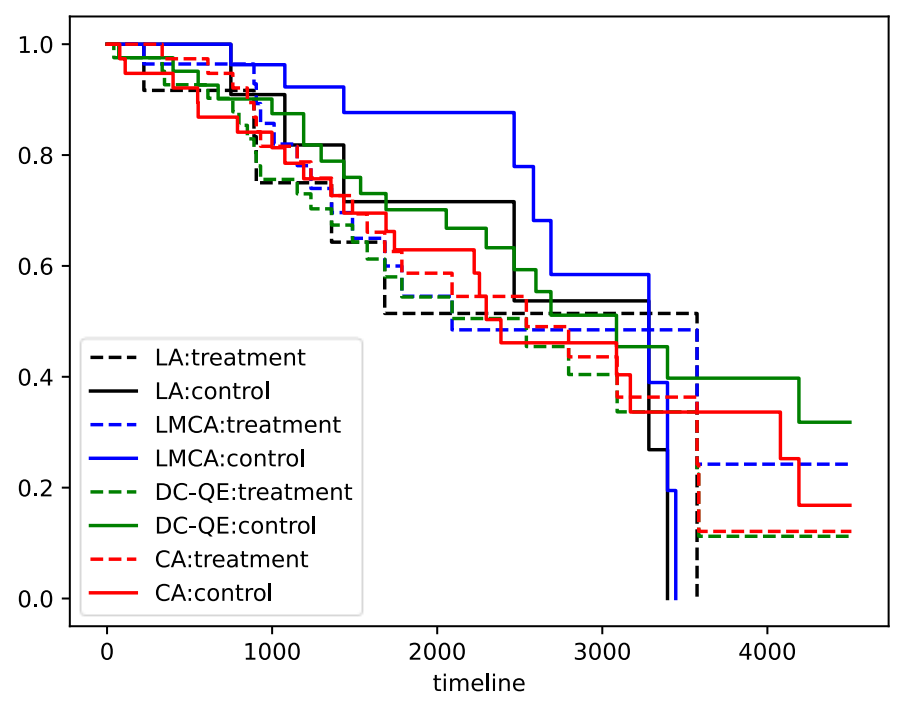

(c) pbc

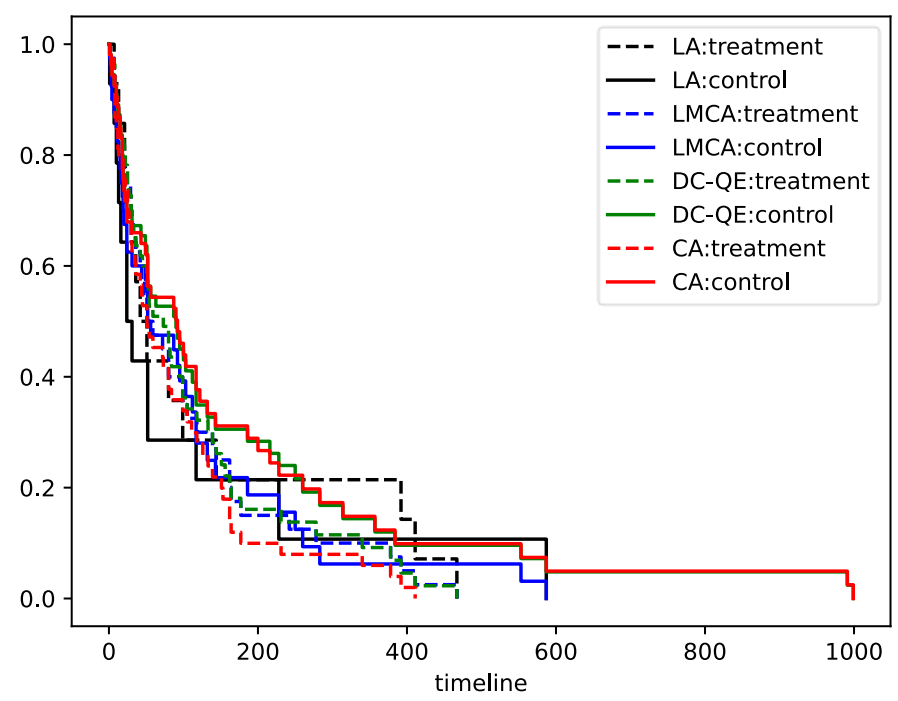

(d) veteran

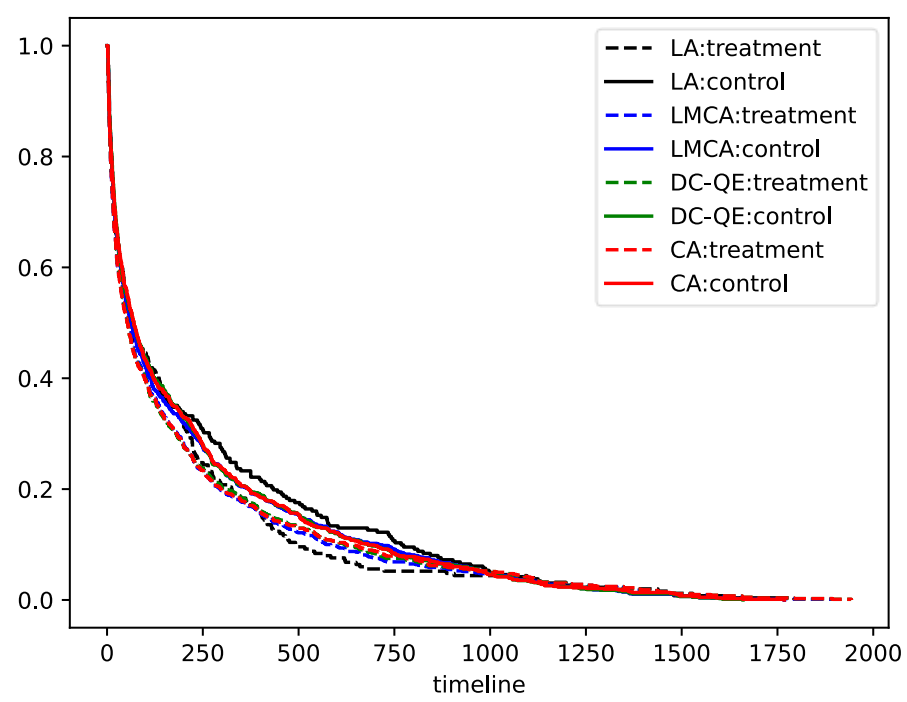

(e) rhc

**Figure 2** Survival curves in Experiment II. (a) to (e) plot the results of survival curves for each dataset. The X-axis represents time, and the Y-axis represents survival probability. In the legend,

solid lines represent the results of the treatment group, and dashed lines represent the results of the control group for each comparison method.

## Discussion

In this study, we propose a novel method for estimating covariate-balanced survival curve in distributed data environments while preserving privacy by extending the DC-QE to survival analysis. In synthetic data experiments, the proposed method significantly improved covariate balance and inconsistency metrics compared to LA and LMCA. DC-QE (W-clb) estimated survival curves closer to the centralized analysis with larger sample sizes. The proposed method most faithfully reproduced both the RMST difference obtained under centralized PSM and the true RMST difference defined in the data generation process. In contrast, LA and LMCA exhibited much larger deviations. Furthermore, experiments using open datasets confirmed the effectiveness of the proposed method across multiple datasets with varying sample sizes and numbers of covariates. The proposed method generated survival curves most similar to the central analysis results. While the difference of RMST difference relative to CA fluctuated across datasets and was occasionally slightly smaller for LMCA, DC-QE simultaneously improved covariate balance and survival curve proximity, achieving better performance than LA.

Unlike simply applying parametric or semiparametric models, such as the Cox model, to distributed environments, the survival analysis approach proposed in this study uniquely combines PSM with the Kaplan–Meier estimator. Nonparametric methods have the advantage of flexibly estimating survival curves based on observed data without requiring prior assumptions about the form of survival time distribution. This flexibility allows them to mitigate bias more effectively when confounders have complex effects. Thus, even in observational studies in which confounding structures are complicated or the number of covariates is large, the proposed method offers the important advantage of accurately estimating survival curves without being constrained by strong model assumptions.

Because DC-QE estimates propensity scores through an integrated framework, even in distributed environments, it facilitates an improved covariate balance between the treatment and control groups. Moreover, although the range of covariate information available for matching may vary depending on the data distribution setting, the proposed method leverages the anchor data along with dimensionality reduction and reconstruction mechanisms, effectively enabling the use of a broader range of covariate information. Consequently, even when individual institutions cannot secure enough samples or covariates on their own, reliable estimation can be achieved by performing matching based on globally integrated information.

We adopted PSM as a covariate balance adjustment method. However, other adjustment techniques based on propensity scores are also available, such as inverse probability weighting and stratification. A major advantage of DC-QE is its flexibility; once the propensity scores are estimated, the framework can be extended to incorporate alternative adjustment methods, such as weighting or stratification, within the context of survival analysis. In practice, it may be beneficial to consider methods other than matching, depending on the characteristics of the survival data. Combining DC-QE with more robust covariate adjustment techniques can further enhance its performance and broaden its applicability.

Several limitations must also be acknowledged. First, even centralized PSM analysis (CA) in simulations retains non-negligible bias relative to the true RMST-based estimand, meaning that all comparative methods—including our distributed approach—are evaluated against the Kaplan–Meier estimator under finite sample size and right censoring, a "realistic but imperfect benchmark." However, since CA, which aggregates all individual-level data to perform PSM, represents the most natural practical analysis procedure and can be considered the gold standard that distributed analysis should approximate, this study adopted CA as the primary comparison method. From this perspective, it is more important to assess how well DC-QE reproduces the same estimand as CA and suppresses additional errors arising from data partitioning; as a supplementary check, the simulation also reports the difference of RMST difference relative to the true survival function. Second, RMST-related metrics in real data are influenced by differences in observation design across datasets, such as follow-up duration and censoring rates. This reflects the fact that RMST captures the explicit quantity "mean survival time up to time $\tau$," and the reliability with which this quantity can be estimated naturally varies across datasets. Therefore, in this study we interpret $D^{\mathrm{CA}}(\tau)$ and $D^{\mathrm{true}}(\tau)$ not as standalone performance metrics, but as one component within a multifaceted evaluation alongside MASMD, survival curve gap, and discrepancy from CA.

Future research directions include extending the framework to handle cases where the treatment variable $Z$ takes multiple values or where time-dependent covariates are involved, and integrating the framework with existing techniques such as differential privacy to establish stronger privacy guarantees. In the current implementation, propensity scores are estimated using logistic regression. Incorporating more flexible machine learning-based propensity score estimation methods could further enhance the method's ability to handle nonlinear confounding structures.

# Conclusion

In conclusion, the proposed method, which extends DC-QE to survival analysis, is shown to significantly improve confounder adjustment compared to LA and achieve covariate-balanced KM curve estimates that closely approximate those obtained by central analysis. Our approach enables a more accurate estimation even in situations where data sharing between facilities or institutions is challenging, making it highly applicable to medical and public health research, which requires balancing privacy preservation with analytical accuracy. In future work, we plan to enhance the practical utility of this method by applying and extending it to more diverse environments and analytical objectives, such as real-world clinical settings and regional public health data.

## Acknowledgments

The authors would like to thank Editage (www.editage.jp) for English-language editing.

## Funding declaration

This work was supported in part by the Japan Society for the Promotion of Science (JSPS), Japan Grants-in-Aid for Scientific Research (Nos. JP22K19767, JP23K22166), Japan Science and Technology Agency (JST) (No. JPMJPF2017) and JST SPRING (No. JPMJSP2124).

## Author contributions

A. T., Y. K., T. N., and Y. O. designed the study. A. T. conducted simulation experiments and data analysis. A. T. and Y. K. wrote the manuscript. Y. K., A. I., and T. S. constructed a DC analysis model tailored to the current problem set and interpreted the results mathematically. All the authors have read and approved the final version of the manuscript.

## Data availability

All data used in this study are either reproducible synthetic data or publicly available data. Synthetic data can be generated using the methods and parameter settings described in the Methods section. The real-world benchmark datasets—lung, veteran, and pbc from the survival package and rhc from the Hmisc package—are all open access and can be obtained from R packages. The datasets used and analysed during the current study are available from the corresponding author on reasonable request.

## Figure legends

**Figure 1** Outline of the proposed method. This method obtains privacy-preserving survival analysis for distributed data without iterative cross-sites communications.

**Figure 2** Survival curves in Experiment II. (a) to (e) plot the results of survival curves for each dataset. The X-axis represents time, and the Y-axis represents survival probability. In the legend, solid lines represent the results of the treatment group, and dashed lines represent the results of the control group for each comparison method.